\documentclass[useAMS,usenatbib]{mn2e}
\usepackage{graphicx,color,afterpage}
\usepackage[total={17.8cm,24.0cm},centering]{geometry}
\def\atlas{{{ATLAS}}$^{\rm 3D}$}
\def\kms{km s$^{-1}$}
\def\msun{M$_{\odot}$}
\def\arcsec{$^{\prime \prime}$}
\definecolor{Mygrey}{gray}{0.75}

\newcommand{\gtsimeq}{\raisebox{-0.6ex}{$\,\stackrel{\raisebox{-.2ex}{$\textstyle >$}}{\sim}\,$}}
\newcommand{\farc}{\mbox{\ensuremath{.\!\!^{\prime\prime}}}}
\title[A FOM for molecular gas SMBH mass measurements]{A figure of merit for black hole mass measurements with molecular gas} 
\author[Timothy A. Davis]{Timothy A. Davis$^{1}$\thanks{E-mail:\texttt{tdavis@eso.org}}
 \vspace{0.4cm}\\
\parbox{\textwidth}{$^{1}$European Southern Observatory, Karl-Schwarzschild-Str. 2, 85748 Garching, Germany\\
}}
\begin{document}
\date{Accepted 2014 June 09. Received 2014 June 09; in original form 2014 March 31}
\pagerange{\pageref{firstpage}--\pageref{lastpage}} \pubyear{2012}
\maketitle
\label{firstpage}
\begin{abstract}
In this work we discuss the technique of using molecular gas kinematics (or the kinematics of any dynamically cold tracer) to estimate black hole masses. We present a figure of merit that will be useful in defining future observational campaigns, and discuss its implications. We show that, in principle, one can estimate black-hole masses using data that only resolve scales $\approx$2 times the formal black hole sphere of influence, and confirm this by reanalysing lower resolution observations of the molecular gas around the black hole in NGC~4526. We go on to discuss the effect that angular resolution, velocity resolution and the depth of the galaxies potential have on the ability to estimate black hole masses, and conclude by discussing prospects for the future. Once ALMA is fully operational, we find that over 10$^5$ local galaxies with massive black holes will be observable, and that given sufficient surface brightness sensitivity one could measure the mass of a $\gtsimeq$4$\times$10$^8$ \msun\ black hole at any redshift. 
\end{abstract}
\section{Introduction}

Over the last twenty years it has become clear that almost all galaxies possess a supermassive massive black hole (SMBH) at their centre. Despite being many thousands of times less massive than the galaxies they live in, and a billion times smaller in size, we now believe that these black holes have major effects on their host systems. 

The empirical relations that have been discovered between galaxy properties and black hole masses \citep[e.g.][]{1998AJ....115.2285M,2000ApJ...539L..13G,2000ApJ...539L...9F,2001ApJ...547..140M,2001ApJ...563L..11G,2003ApJ...589L..21M,2009ApJ...698..198G}, suggest that galaxies and black-holes may co-evolve in a self-regulating manner \citep[e.g.][]{1998A&A...331L...1S,2003ApJ...596L..27K}. 

SMBH demographics also reveal that different galaxy types have different evolutionary histories.
For instance, black-hole masses may not correlate with the properties of galactic disks (that are thought to mainly grow through secular processes), but correlate strongly with the properties of bulges \citep{2013ARA&A..51..511K}. Spiral galaxies, that on average have a quieter formation history, also seem to have smaller black-holes than early-type systems even for the same bulge velocity dispersion \citep{2013ApJ...764..184M}.

Obtaining accurately measured black hole masses is thus a powerful way to probe the processes involved in galaxy evolution. Such observations are challenging, however, requiring high angular resolution observations with reasonable spectral resolution.
Three main methods have been used to date to directly measure black-hole masses: observations of stellar kinematics (mainly in early-type galaxies; e.g. \citealt{1988ApJ...324..701D,1988ApJ...325..128K,2003ApJ...583...92G}), ionised-gas kinematics (in spiral and some early-type galaxies; e.g. \citealt{1996ApJ...470..444F,2001ApJ...550...65S,2001ApJ...555..685B,2002PASP..114..137H}) and the kinematics of nuclear masers (in rare objects that have suitable central masing disks; e.g. \citealt{1995Natur.373..127M,1997ApJ...481L..23G,1999JApA...20..165M,2005ARA&A..43..625L}).

Recently, \citet[][hereafter D13]{2013Natur.494..328D} introduced a new tracer which can be used to estimate black hole masses; using millimetre interferometry to resolve the kinematics of molecular clouds in the centre of the lenticular galaxy NGC~4526. 
This galaxy lies at 16.5 Mpc, has an inclination of 79$^{\circ}$  and was found to have a SMBH mass of $\approx$4$\times$10$^8$ \msun. As highlighted in D13, molecular gas as a tracer has various useful quantities; in principle it is possible in any galaxy type, and the high angular resolutions achievable by new (sub)-millimetre interferometers (e.g. the Atacama Large Millimetre/submillimetre Array; ALMA) mean it may be possible to probe larger volumes of the universe than ever before.  

In this paper we consider the strengths and weaknesses of using molecular gas kinematics to estimate black hole masses, and introduce a figure of merit that should be useful when planning future observational campaigns. We briefly describe the method, deriving this figure of merit in Section \ref{FOMsection}, and go on to highlight its possible applications using observational data in Section \ref{applications}. We then conclude and reflect on future prospects in Sections \ref{futuresamples} and \ref{conclude}. Throughout this work we use the the $M_{\rm BH}$-$\sigma_{*}$ relations of \cite{2013ApJ...764..184M}, and a cosmology where $\Omega_m$=0.3, $\Omega_\lambda$=0.7 and H$_0$=71 \kms Mpc$^{-1}$, unless otherwise stated.

\section{Technique and Figure of Merit}

{The procedure by which one can use molecular gas to estimate SMBH masses builds on the established method used in ionised gas measurements.
Observations of the Keplerian kinematics of a molecular disk are used estimate the total (luminous plus dark) gravitational potential of the target object. A de-projected mass model of the galaxy in question (usually constructed from high resolution near-infrared imaging) is then used to estimate the kinematic signal expected from just
the luminous stellar mass of the bulge regions. The stellar mass-to-light ratio (M/L) can be fitted as a free parameter using this molecular gas data, or constrained through other kinematic or stellar population studies at larger radii. In the absence of very strange dark-matter profiles, the kinematic effect of the dark matter halo varies smoothly in the inner parts of galaxies, and is thus usually thought to be included with this fitted M/L term.}

{ In galaxies where the molecular gas density in the inner parts is an important contribution to the total mass density, the high resolution gas image also obtained from the interferometer can be used to include its contribution to the potential (e.g. \citealt{2011ApJ...735...88A}). This correction for the potential of the molecular material was not done in D13, as the total molecular mass within the inner 80pc is only $\approx$2$\times$10$^6$ M$_{\odot}$, at least two orders of magnitude smaller than the stellar mass in the same region. When generalising this technique to more gas rich galaxies, however, such a correction may be important. }

{Once a suitable mass model has been created, forward modelling is used to predict the observed kinematics expected given the luminous matter alone. The difference between the kinematics from the luminous mass model and the observed kinematics from the molecular disk can then be estimated. If a significant difference is found, additional models can be created including the mass of a central dark object, and these can then be fitted to the data to determine a best fit SMBH mass. }

\subsection{Figure of Merit}
\label{FOMsection}
In order to detect the kinematic signature of a SMBH using molecular gas as your tracer, one must observe molecular emission at higher velocities than predicted from the potential of the luminous mass alone. Let us define v$_{\rm gal}$($r$) {as the speed a test particle (i.e. a molecular cloud) would have in a circular orbit in the equatorial plane of an edge on galaxy at radius $r$, given the potential of the luminous mass alone} (see Sections \ref{chansizeres} and \ref{futuresamples} for a further discussion on the importance of accurately determining this quantity). {We here only consider the case of dynamically cold gas in a flat disk (or ring) rotating at the circular velocity. In cases where the gas is not dynamically cold, warped, or inflowing/outflowing this analysis is not formally valid (we discuss these cases further below).}

We then define v$_{\rm obs}$(r) as the observed velocity of a given gas parcel at a radius $r$, including the effect of a SMBH. To claim a detection of the SMBH signal at confidence level $\alpha$ (i.e. $\alpha$=3 is a 3$\sigma$ detection at radius $r$), the projected velocity difference at this radius $r$ (usually defined as the closest resolvable distance to the centre of the galaxy of interest, that is determined by the beamsize of the telescope) must be at least $\alpha$ times larger than the associated error ($\delta v$) in both the observational quantities and the modelling. If the galaxy is seen at an inclination to our line of sight $i$, then

\begin{equation}
v_{\rm obs}|_{r} - v_{\rm gal}|_{r}  \sin{i} > \alpha \, \delta v.
\label{baseeq}
\end{equation}

\noindent Assuming the gas is on purely circular orbits in a flat disk that shares the same inclination ($i$) as the galaxy, then

\begin{equation}
v_{\rm obs}(r) = \sqrt{[v_{\rm gal}(r)^2 - \phi_{\rm BH}(r)]}  \sin{i},
\end{equation}

\noindent where $\phi_{\rm BH}$ is the gravitational potential of the SMBH (${\frac{-GM_{\rm BH}}{r}}$), with G being the gravitational constant, and $M_{\rm BH}$ being the black hole mass.
Substituting this into Equation \ref{baseeq} we obtain

\begin{equation}
\sqrt{v_{\rm gal}(r)^2 - \phi_{\rm BH}(R)} - v_{\rm gal}(r)> \frac{ \alpha \, \delta v}{\sin{i}}.
\end{equation}

Through a simple rearrangement, this becomes the basic equation for our figure of merit ($\Gamma_{\rm FOM}$)
\begin{equation}
\Gamma_{\rm FOM} = \frac{\sqrt{[v_{\rm gal}(r)^2 - \phi_{\rm BH}(r)]} - v_{\rm gal}(r)} { \alpha \, \delta v} \sin{i},
\label{base_eqnofmerit}
\end{equation}

\noindent which is equal to one for a detection of the SMBH signature at a confidence level $\alpha$. 

Assuming that the point at which one wishes to estimate the black hole mass is the closest resolvable distance from the centre (in the limit of very good sampling of the spatial PSF) then one can further redefine that in parsecs $r$=4.84$\theta D$ where $\theta$ is the telescope beam size in arcseconds and D is the distance to the galaxy in megaparsecs (the factor 4.84 comes from the definition of a parsec).

 Several useful formulae follow directly from the above definition of the figure of merit (Equation \ref{base_eqnofmerit}), and are presented in Section \ref{usefulform} for the readers convenience. 

 \subsection{Error terms}
 The exact form of the error term $\delta$v in the above figure of merit (Equation \ref{base_eqnofmerit}) depends on the individual situation. Several error terms are almost always present, but more may be required in specific cases. 
 
 The first error term that must be included for radio interferometer data is that induced by the channellisation of the data ($\delta v|_{\rm chan}$). 
  For a dynamically cold tracer like molecular gas, the intrinsic line width of 5-10 \kms {(caused by the combination of inter-cloud velocity dispersion with the small thermal line width of individual molecular clouds)} is usually smaller than the channel width. In this limit one is unable to tell the true velocity of a parcel of gas to better than half the channel width. If the model data and the real data have been treated in the same way then:
 
 \begin{equation}
\delta v|_{\rm chan} = \sqrt{2\left(\frac{{\rm CW}}{2}\right)^2} = \sqrt{0.5}{\rm CW}
\end{equation}

If the line width is instead well sampled then one can determine the velocity centroid of the emission line more accurately, and this term becomes a function of both the channel width and the signal-to-noise ratio.

In addition the velocity estimated from a model of the luminous matter in the system is likely to have an associated error ($\delta$V$|_{\rm gal}$). The accuracy of your mass model is thus key in order to obtain an accurate estimate of a SMBH mass. In individual situations one may also want to add error terms taking into account observational error on the derived inclinations, deviations from circular motions (e.g. inflow, outflow, the presence of spiral structure), etc. All these error terms should be added in quadrature:
 \begin{equation}
\delta v|_{\rm tot} =  \sqrt{ 0.5({\rm CW})^2 + \delta v|^2_{\rm gal} + ...}.
\label{erroreq}
 \end{equation}
 
 \noindent We shall assume that only channellisation and mass model errors are important in what follows, but caution that other error terms may be needed in more general cases.

  \subsection{Useful formulae}
  \label{usefulform}
 In this section we present useful formulae that follow from the definition of the figure of merit (Equation \ref{base_eqnofmerit}).
 Firstly, the smallest black hole one can reliably detect (i.e. $\Gamma_{\rm FOM}$=1) at confidence $\alpha$, at a distance $D$, with a resolution of $r$ parsecs, and in an object with an inclination $i$ and a circular velocity caused by luminous matter of v$_{\rm gal}$ (at position $r$) is
 
  \begin{equation}
M_{\rm BH}|_{\rm min}=  \frac{r \alpha \, \delta v}{G} \left( \frac{ \alpha \, \delta v}{\sin^2{i}} +  \frac{2v_{\rm gal}}{\sin{i}}\right).
 \end{equation}

The furthest away one can detect a black hole with mass $M_{\rm BH}$ at a confidence $\alpha$, with a beam of $\theta$ arcseconds, with a circular velocity caused by luminous matter of V$_{\rm gal}(\theta)$ is

  \begin{equation}
D|_{\rm max}=  \frac{GM_{\rm BH}}{4.84 \theta \alpha \, \delta v \left( \alpha \, \delta v \sin^{-2}{i} +  2v_{\rm gal} \sin^{-1}{i}\right)},
\end{equation}

\noindent and equivalently the angular resolution (in parsecs) that is required to detect a black hole with mass $M_{\rm BH}$ at a confidence $\alpha$ in an object with a circular velocity caused by luminous matter of V$_{\rm gal}(\theta)$ is

  \begin{equation}
r|_{\rm max}=  \frac{GM_{\rm BH}}{ \alpha \, \delta v \left( \alpha \, \delta v \sin^{-2}{i} +  2v_{\rm gal}(r) \sin^{-1}{i}\right)}.
\label{soireplace}
\end{equation}

 It should be remembered here that generally $v_{\rm gal}$ is a function of $r$ and so this quantity may be best estimated by evaluating $\Gamma_{\rm FOM}$ as a function of radius, and determining where $\Gamma_{\rm FOM}$=1. 
 
Less usefully one can also estimate the maximum circular velocity caused by luminous matter possible in a galaxy to still allow a detection of an SMBH signature (with symbols as defined before):

\begin{equation}
 v_{\rm gal}|_{\rm max} = 0.5 \left[\left(\frac{-\phi(r) \sin{i}}{ \alpha \, \delta v}\right) - \left(\frac{\alpha \, \delta v}{\sin{i}}\right)\right]. 
\end{equation}

By substituting in Equation \ref{erroreq} into Equation \ref{base_eqnofmerit} it is also possible to show that that the maximum channel width to detect a SMBH with mass $M_{\rm BH}$ at a confidence $\alpha$ (with symbols as previously defined) is 

\begin{eqnarray}
{CW}|_{\rm max} = \sqrt{2\left[\left({{v_{\rm gal}^2 - {\phi(r)}} - v_{\rm gal}} \right)^{0.5}{ \frac{\sin{i}}{\alpha}}\right]^2 - 2dv|_{\rm errs}^2 },
\label{fom_cw}
\end{eqnarray}

\noindent where $dV|_{errs}^2$ is the quadratic sum of all the error terms excluding the channel width (as shown in Equation \ref{erroreq}).

\section{Applications}
\label{applications}
\subsection{Modifications to the sphere of influence criteria}
\label{changesoi}

The size of the region in which the gravitational potential of a SMBH dominates the gravitational potential of the host galaxy is called the sphere of influence (SOI).
It is possible in some cases to estimate black-hole masses outside the formal SOI using ionised gas and stellar methods \citep[e.g.][]{2003ApJ...583...92G,2006A&A...460..449W,2010AIPC.1240..211C}, however this criteria is often used in practise to roughly determine which SMBHs are observable \citep[e.g.][]{2005SSRv..116..523F}.
The SOI is usually calculated as 

\begin{equation}
r_{\rm SOI} =  \frac{GM_{\rm BH}}{\sigma_{*}^2},
\label{classicsoi}
\end{equation}
\noindent where $\sigma_{*}$ is the stellar velocity dispersion, and the other symbols are as defined above. 
{A different formalism is also occasionally used, where the sphere of influence is defined as the region enclosing a total mass twice that of the SMBH mass (i.e. M(r$< r_{\rm SOI,m}$) = 2M$_{\rm BH}$). These two sphere of influence definitions are equal in the case of an isothermal sphere, but differ in real objects.}
It has been claimed that the application of these formulae to determine which galaxies are likely to have measurable black-holes could lead to biases in the derived SMBH-galaxy relations (e.g. \citealt{2010ApJ...711L.108B}), and so any technique that could push below this limit would be useful. 

Equation \ref{soireplace} suggests that the enhancement of the circular velocity due to the black-hole is measurable further out than the classical SOI criteria would indicate.
Thus we argue that classical SOI criteria needs to be modified when one is dealing with molecular gas measurements, or indeed any technique where the majority of the signal is seen in velocity rather than velocity dispersion (see \citealt{2013degn.book.....M} for a similar discussion). 
{We are clearly in such a velocity dominated regime here. For instance, taking the case of NGC~4526, the gas velocity dispersion ($\sigma_{\rm g}$) is consistent with being less than 10 \kms\ throughout the molecular disk, while the velocity gradient across a single interferometer beam element can be as high as 240 \kms. As the generic signature one requires to detect an SMBH is a cusp in v$_{\rm obs}^2$+$\sigma_{\rm g}^2$ \citep[e.g.][]{2013degn.book.....M}, it is clear that for molecular gas the velocity component dwarfs the contribution from the gas velocity dispersion.}

\begin{figure}
\begin{center}
\includegraphics[width=0.49\textwidth,angle=0,clip,trim=0.0cm 0.0cm 0cm 0.0cm]{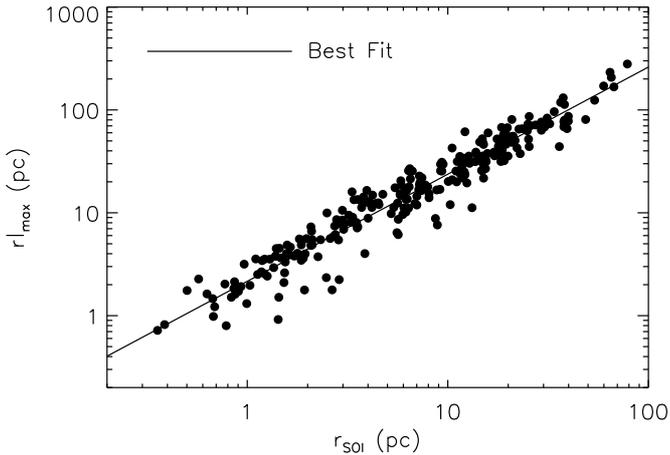}
 \end{center}
\caption{Sphere of influence (derived with Equation \ref{classicsoi}) for all the early-type galaxies from the \atlas\ survey, plotted against r$|_{\rm max}$ (the size of the region in which you can detect the expected SMBH at $\alpha$=5 using molecular gas) as defined using our figure of merit. Overplotted in black is the best-fit line, as specified in the text. }
\label{rsoivsrgamma}
 \end{figure}

In order to demonstrate the typical difference between $r_{\rm SOI}$ from Equation \ref{classicsoi} and $\theta|_{\rm max}$ from Equation \ref{soireplace} we retrieved velocity dispersions, inclinations, distances and circular velocity profiles for every galaxy in the \atlas\ survey \citep{2011MNRAS.413..813C}. Using the {early-type galaxy} M$_{\rm BH}$-$\sigma$ correlation of \cite{2013ApJ...764..184M} we estimate the black-hole mass of each object, and thus its SOI. We then also calculated $r|_{\rm max}$ (see Equation \ref{soireplace}, as the location where our figure of merit suggests that the enhancement in the circular velocity would be 5$\times$ our velocity errors (i.e. where $\Gamma_{\rm FOM}(\alpha=5$) is equal to one). We plot $r_{\rm SOI}$ against $r|_{\rm max}$ in Figure \ref{rsoivsrgamma} for each of the \atlas\ galaxies (black points). 

There is a correlation between these two measures (as is to be expected give that they both are a measure of the gravitational effect of the same mass SMBH in the same potential), but on average $r|_{\rm max}$ is $\approx$2 times larger than $r_{\rm SOI}$. We used a simple least-squares linear regression to determine a best fit relation between these quantities, finding $r|_{\rm max}$ = 1.94$\pm$0.07 $R_{\rm SOI}^{1.02\pm0.02}$  (as shown by the black line in Figure \ref{rsoivsrgamma}).  The scatter around this relation is 0.15 dex for our sample galaxies. {As shown in Figure 2.4 in \cite{2013degn.book.....M}, the area enclosing twice the mass of the SMBH ($R_{\rm SOI,m}$) is approximately 2$R_{\rm SOI}$ for the typical \cite{1948AnAp...11..247D} density profile of early-type galaxies, providing a physical explanation for the size of the observed effect. However, our estimation here takes into account the possible sources of error, and thus for different error terms and desired confidence levels (i.e. $\alpha$ $\neq$ 5), the magnitude of this factor will change.}

Our analysis above suggests one can estimate black-hole masses with a resolution $\approx$2 lower than required to satisfy the classical sphere of influence criteria. This clearly has significant implications for the number of accessible black-holes in the local universe (we go on to discuss these prospects further in Section \ref{futuresamples}). In the next section we test the validity of this analysis, using the black hole in NGC~4526 as a test case.

\subsubsection{The SMBH in NGC~4526}
\label{smbh4526}

 \begin{figure*}
\begin{center}
\includegraphics[width=0.48\textwidth,angle=0,clip,trim=0.0cm 0.0cm 0cm 0.0cm]{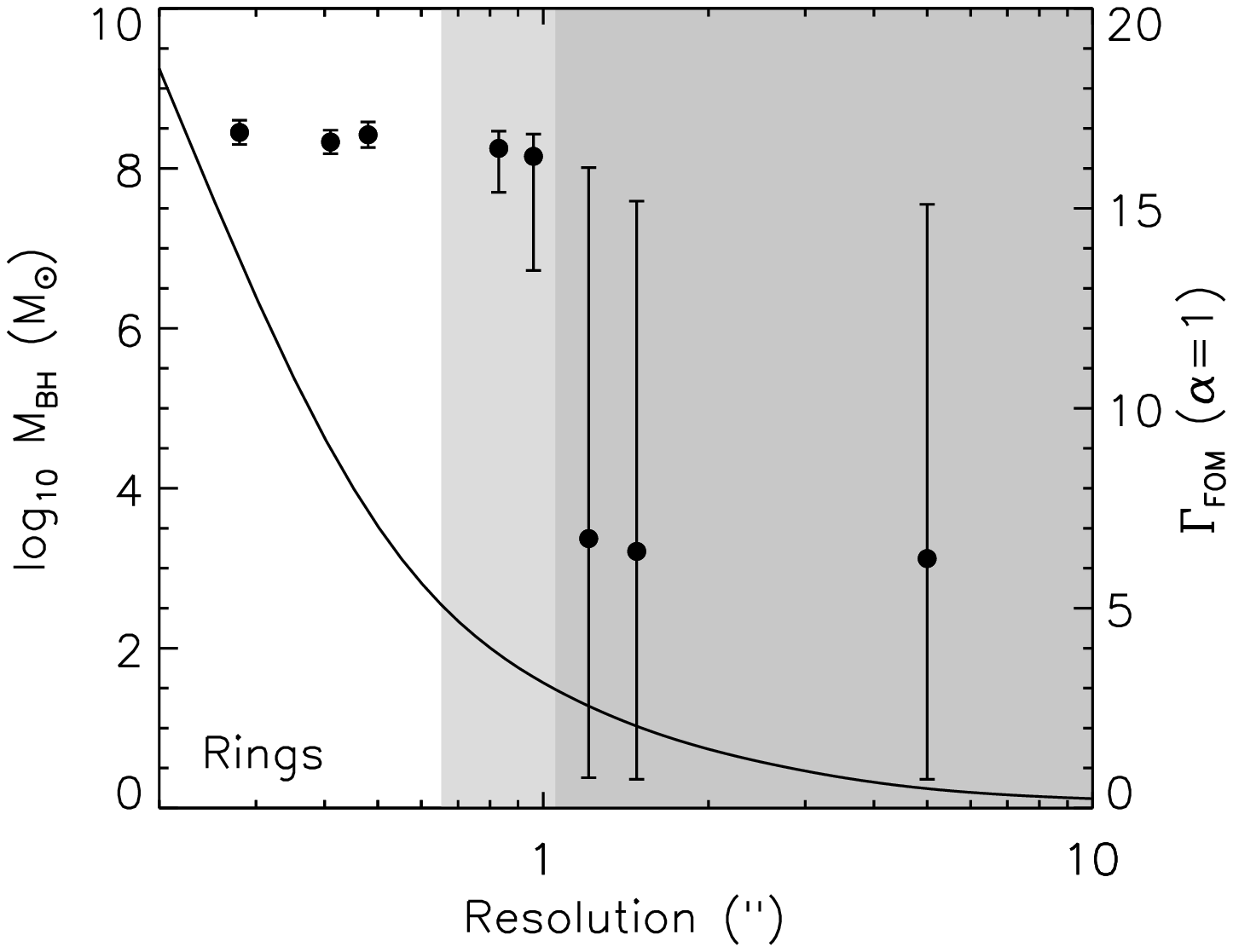} \hspace{0.5cm}
\includegraphics[width=0.48\textwidth,angle=0,clip,trim=0.0cm 0.0cm 0cm 0.0cm]{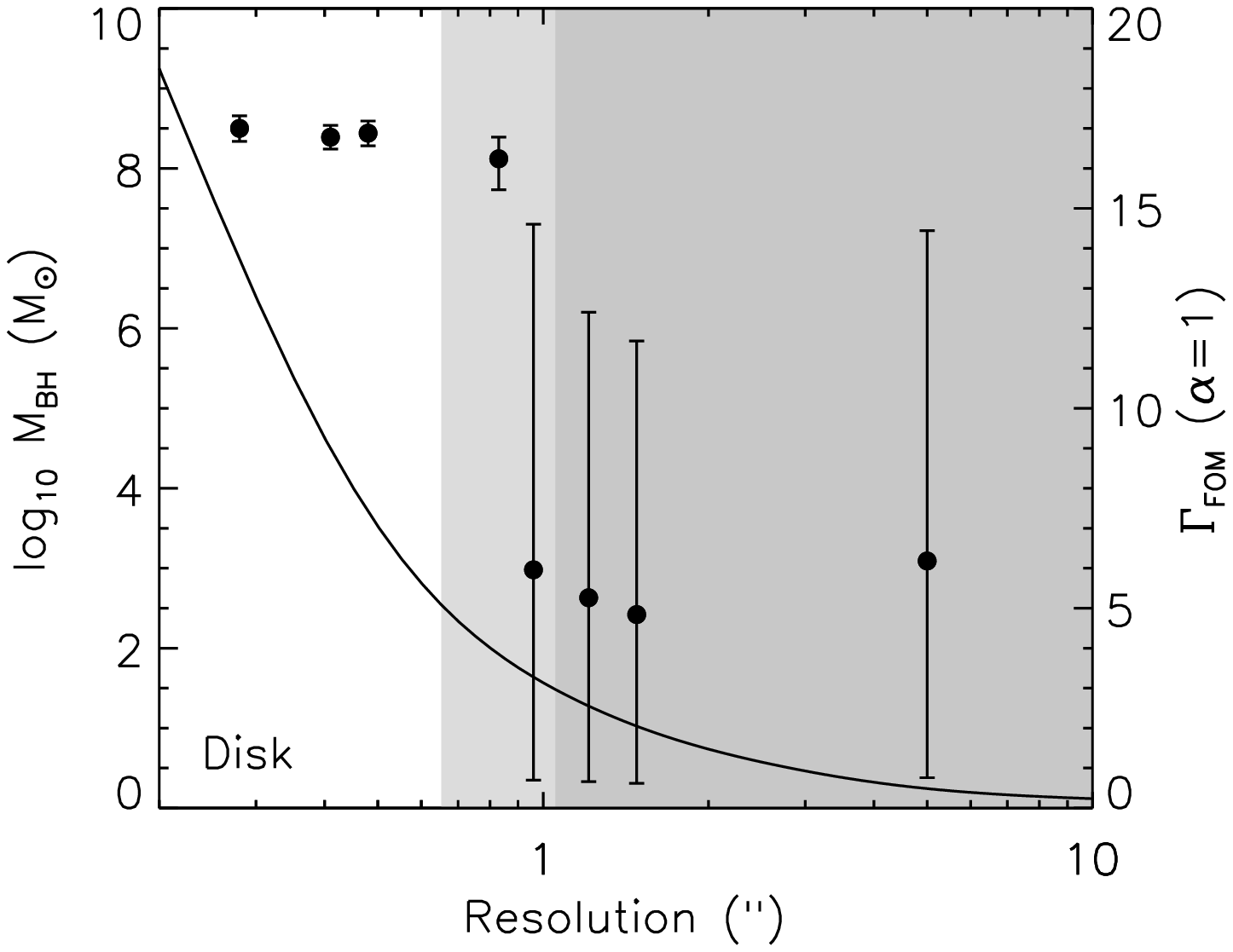}
 \end{center}
\caption{ Mass estimates (with formal 3$\sigma$ error bars) for the massive dark object in NGC~4526, plotted as a function of the resolution of the observational data used. The points at $<$2\arcsec\ are the NGC~4526 CO(2-1) data from CARMA (as presented in D13) imaged to produce different resolutions, while the point at 5\arcsec\ is the BIMA data of \protect \cite{2008ApJ...676..317Y}. Black hole mass estimates are made with the \textsc{KinMS\_mcmc} code and the best fits and errors are determined here from 10$^5$ trials. Error bars for the low resolution points should not be thought of as representative of the full range of allowed solutions. Also plotted is $\Gamma_{\rm FOM}$($\alpha$=1) (black line) as a function of resolution for this galaxy. The area with $\Gamma_{\rm FOM} < 3$ (where we do not expect to be able to detect the black holes influence) is shaded dark grey, and the area with $\Gamma_{\rm FOM} < 5$ is shaded a lighter grey. In the left panel we use a surface brightness profile model with 3 rings, as revealed by the high resolution interferometric imaging. In the right panel we assume this information is not available and fit an exponential disk model for this object (as determined from very low resolution data in \citealt{2013MNRAS.429..534D}).}
\label{bhvsres}
 \end{figure*}

In order to validate our numerical prediction (made in Section \ref{changesoi} above) that it is possible to detect an SMBH using molecular gas as a tracer to good significance even outside the formal SOI of $0\farc25$ (D13) we here perform a test, attempting to detect the SMBH in NGC~4526 using the observational data from D13 re-imaged to yield different angular resolutions.

 The CO(2-1) CARMA observations used in D13 had a beam of 0\farc27$\times$0\farc17 (giving an effective resolution of $0\farc25$ along the major axis), and a velocity resolution of 10 \kms. The circular velocity caused by luminous matter of the galaxy at 0\farc25 is $\approx$150 \kms, and the error on that quantity arising from the mass modelling is $\approx$5 \kms. Thus using our figure of merit formalism, the dark object in the centre of NGC~4526 was detected at $\approx$19 times the error level (i.e $\Gamma_{\rm FOM}$=18.46 with $\alpha$=1). This suggests that indeed we should be able to detect the influence of the black hole out to larger radii. Using Equation \ref{soireplace} with $\alpha$=3 our formalism suggests we should be able to detect the signature of the SMBH and measure its mass out to a radius of $\approx$1\arcsec.

In order to test this, we re-imaged the calibrated visibilities, using different weighting and tapering schemes to produce datasets with resolutions of 0\farc41, 0\farc48, 0\farc83, 0\farc96, 1\farc21 and 1\farc48. In addition we used the low resolution CO(1-0) BIMA data of \cite{2008ApJ...676..317Y} (resolution 5\arcsec). We input this data to a new Markov Chain Monte Carlo (MCMC) code (\textsc{KinMS\_mcmc}) that couples to the KINematic Molecular Simulation (\textsc{KinMS}) routines\footnote{available at http://purl.org/KinMS} presented in \cite{2013MNRAS.429..534D}, and allows us to fit the data and obtain the full bayesian posterior probability distribution for the fitted parameters. This code fits the entire data cube produced by the interferometer, rather than simply the position-velocity diagram (as was done in D13). 

In the left panel of Figure \ref{bhvsres} we used the same parameters (surface brightness profile, total flux, inclination, centre position, etc) as D13, and fit for the black-hole mass (that is given a flat prior in log-space, and allowed to vary between 0 and 10$^{10}$ \msun) and the mass-to-light ratio at $I$-band (that was given a flat prior in linear space, and allowed to vary between 1 and 4 L$_{\odot}$/\msun). The resulting best-fit black-hole masses (marginalised over the M/L) are then plotted as a function of resolution. We also show as a black line the $\Gamma_{\rm FOM}(\alpha=1)$ profile, and shade the areas where $\Gamma_{\rm FOM}<$ 3 and $\Gamma_{\rm FOM}<$ 5. As clearly demonstrated in the left panel of Figure \ref{bhvsres} we are able to estimate the SMBH mass in this galaxy at all resolutions $<1$\arcsec. In the areas where $\Gamma_{\rm FOM}<$3 we are unable to detect signs of a SMBH (the error bars on these points should not be taken as limits on the parameters, as when parameters are unconstrained it is difficult to ensure the MCMC algorithm properly explores all the available parameter space). This test suggests that in principle black-hole mass measurements made with molecular gas may be possible well outside the formal SOI.

The test described above used the observed ring morphology of the gas to fit the data at all resolutions, despite the fact that at lower resolutions the real gas distribution would not be obvious (see \citealt{2008ApJ...676..317Y,2013MNRAS.429..534D,2013MNRAS.432.1796A}). We test if an incorrect assumption about the gas surface brightness profile could cause problems in detecting the signature of a black hole. In the right panel of Figure \ref{bhvsres} we repeat the experiment above, but use an exponential disk as our input surface brightness profile for the models.  We obtain a very similar result to that shown in the left panel, obtaining entirely consistent results for the higher resolution points. We however fail to detect the SMBH at a resolution of 0\farc96. We thus recommend using $\Gamma_{\rm FOM}(\alpha=5)$ as a conservative threshold to aim for when planning black-hole mass measurements, to ensure a good detection even without \textit{a priori} knowledge of the gas distribution. 

{As discussed above, the FOM analysis presented in this paper only holds in cases like NGC~4526, where the molecular gas appears to be in dynamically cold thin disks/rings, and only circular motions are important. Lower resolution observations suggest that such dynamically cold disks are common in molecular gas hosting ETGs and flocculent spirals \citep[e.g.][]{2002PASP..114..137H,2013MNRAS.429..534D}, but galaxies with stronger spiral patterns may have contributions from non-circular streaming motions \citep[e.g.][]{2009ApJ...692.1623H}.}  Having a higher angular resolution than suggested by our $\Gamma_{\rm FOM}(\alpha=5)$ criteria is sensible in cases where one expects non-circular/streaming motions or disk warps to be significant, as having many resolution elements within a source makes it easier to identify and model such phenomena.

\subsection{Minimum channel sizes}
\label{chansizeres}
(Sub-)millimetre interferometers have an advantage over typical optical spectroscopy, in that it is possible to routinely obtain very high spectral resolution ($<$1 \kms)  observations. In principle, simply observing a potential SMBH host galaxy at very high spectral resolution would allow the detection of an SMBH even very far out in the molecular disk. In practice however, the accuracy with which you can create the mass model for the luminous matter in your system (and the possible presence of streaming or non-circular motions in the gas) limits the minimum desirable spectral resolution.

\begin{figure}
\begin{center}
\includegraphics[width=0.49\textwidth,angle=0,clip,trim=0.0cm 0.0cm 0cm 0.0cm]{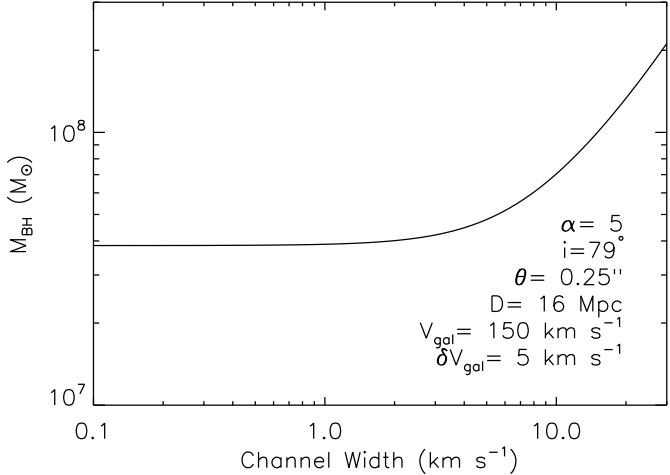}
 \end{center}
\caption{Minimum SMBH mass detectable (at $\alpha$=5) as a function of the channel width using in the observations (assuming the observational parameters are as for NGC~4526; see Section \ref{smbh4526} and legend). }
\label{minchanwidth}
 \end{figure}

Equation \ref{fom_cw} and Figure \ref{minchanwidth} show clearly that one cannot simply decrease the channel width to arbitrarily small values to increase the accuracy of your SMBH measurement. The smallest SMBH mass you can measure begins to asymptote when ones channel width becomes equivalent to the size of these errors in the mass model (and any additional error terms). Thus, for example, for the NGC~4526 case discussed above little can be gained by decreasing the channel width below $\approx$5 \kms.

\subsection{Mass profile}

 \begin{figure}
\begin{center}
\includegraphics[width=0.49\textwidth,angle=0,clip,trim=0.0cm 0.0cm 0cm 0.0cm]{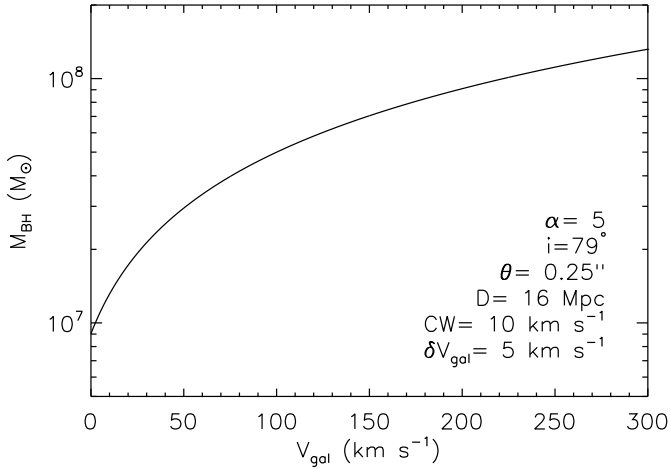}
 \end{center}
\caption{Minimum SMBH mass detectable (at $\alpha$=5) as a function of the galaxy circular velocity (caused by luminous matter) at the beam size using in the observations (assuming the observational parameters are as for NGC~4526; see Section \ref{smbh4526} and legend).}
\label{massprof}
 \end{figure}

As shown in Equation \ref{base_eqnofmerit}, two galaxy properties enter the equation for our figure of merit, the black-hole mass itself, and the circular velocity due to luminous matter at the desired angular resolution. This latter parameter will be different in galaxies that have a different mass profile (e.g. a core or cusp), even if the galaxy itself hosts the same mass black-hole. 

\begin{figure}
\begin{center}
\includegraphics[width=0.49\textwidth,angle=0,clip,trim=0.0cm 0.0cm 0cm 0.0cm]{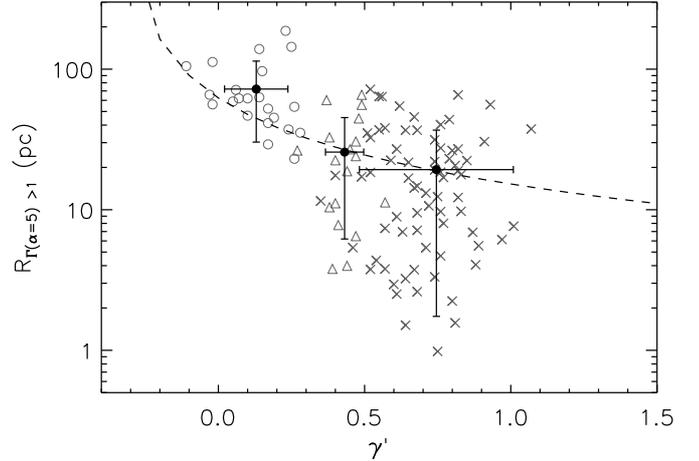}
 \end{center}
\caption{The \atlas\ galaxies are plotted in the $R_{\Gamma(\alpha=5) > 1}$ (the size of the region in which you can detect the expected SMBH at $\alpha$=5) versus slope of the potential $\gamma$' (defined as the logarithmic derivative of the light profile at a radius of 0\farc1) plane. Objects with cores are shown as grey open circles, objects with power law cusps as crosses, and intermediate objects as open triangles, and the mean value for each class is shown as a black point (with error bars equal to the standard deviation of the points). Black-hole masses have been estimated assuming these objects lies on the elliptical galaxy M$_{\rm BH}$-$\sigma$ relation, and $\gamma$' is taken from \protect \cite{2013MNRAS.433.2812K}. The dashed line shows the relation expected for average galaxy parameters, as explained in the text.}
\label{massprofgamma}
 \end{figure}

Figures \ref{massprof} and  \ref{massprofgamma} show the importance that the mass profile of the galaxy plays in setting the limiting SMBH mass one can detect. 
Figure \ref{massprof} shows the minimum mass SMBH one can detect (at $\alpha$=5) as a function of the circular velocity at 20pc (0\farc25 in this example - applicable for Virgo cluster objects).
It is clear that it will be possible to detect a smaller SMBH if the circular velocity at the telescope resolution is low.

 In Figure \ref{massprofgamma} we show an observational example, using the \atlas\ galaxies (as in Figure \ref{rsoivsrgamma}). We here plot $R_{\Gamma(\alpha=5) > 1}$, the expected size of the region in which you can detect the signature of a SMBH (that lies on the M$_{\rm BH}$-$\sigma$ relation) at $\alpha$=5 as defined in Section \ref{changesoi}, plotted against the slope of the inner light profile of these objects $\gamma$' (as measured in \citealt{2013MNRAS.433.2812K}). Objects with cores are shown as grey open circles, objects with power law cusps as crosses, and intermediate objects as open triangles. The real galaxies show significant scatter, but the mean values for the 3 classes (shown as solid points with error bars) show a negative correlation. 
 
 We also overplot a theoretical line, derived using Equation \ref{soireplace} using average values of the parameters for the \atlas\ sample (a black hole mass of 7.5$\times10^7$\msun, a distance of 23 Mpc). We estimate the mean V$_{\rm gal}$ as a function of $\gamma$' from the \atlas\ data, and set the other parameters to the same values as used in Figures \ref{minchanwidth}. We show the resulting prediction as a dashed line, and find that it does a reasonable job of reproducing the mean trend. 

Overall we conclude that it is significantly easier to detect a given size SMBH if the galaxy potential is shallow (e.g objects with cores) than if it is steep (e.g. objects with power law or cuspy profiles). Stellar black hole mass measurements are often easier in cuspy objects (as long as an unresolved nuclear star-cluster is not present) {due to the greater concentration of tracer stars in the inner regions, highlighting the complementarity of these methods.}

\section{Discussion}
\label{futuresamples}
 \begin{figure}
\begin{center}
\includegraphics[width=0.49\textwidth,angle=0,clip,trim=0.0cm 0.0cm 0cm 0.0cm]{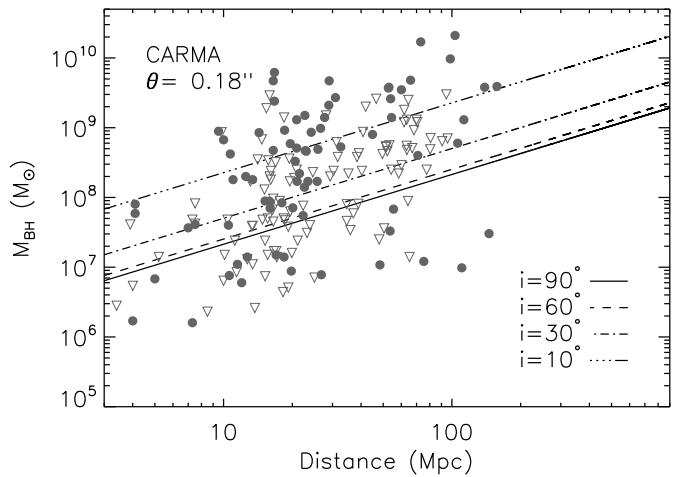}
 \end{center}
\caption{Minimum SMBH mass detectable (at $\alpha$=5) as a function of luminosity distance by CARMA (highest angular resolution at 230 GHz of 0\farc18), in galaxies with V$_{\rm gal}$=100$\pm$5 \kms\ and with a channel width of 10 \kms. Dashed lines (as indicated in the legend) show the limits reachable for galaxies at different inclinations. The grey solid points in both plots are the galaxies on which the M-$\sigma$ relation is currently based (from the catalogue of \citealt{2013ApJ...764..184M}), and the grey triangles are the SMBH mass upper limits from \protect \cite{2009ApJ...692..856B}.}
\label{almacarma}
 \end{figure}
 
  \begin{figure}
\begin{center}
\includegraphics[width=0.49\textwidth,angle=0,clip,trim=0.0cm 0.0cm 0cm 0.0cm]{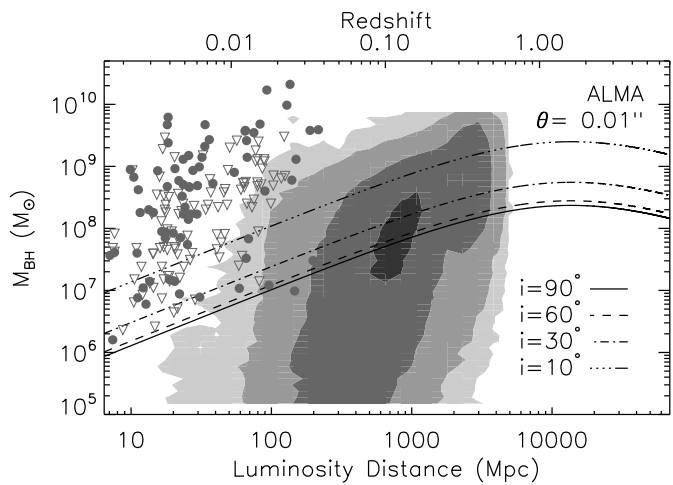}
 \end{center}
\caption{As Figure \protect \ref{almacarma} but showing the limits reachable by full ALMA at CO(3-2) (highest angular resolution at 345 GHz of 11 mas). Using higher frequency bands will further reduce the minimum detectable SMBH mass. We also show as shaded contours the distribution of galaxies with measured velocity dispersions from the SDSS DR7 database \citep{2009ApJS..182..543A}, with M$_{\rm BH}$ estimated from the M$_{\rm BH}$-$\sigma$ relation. }
\label{almacarma2}
 \end{figure}

If using molecular gas as a tracer is to be a useful addition to the toolkit of techniques to estimate black-hole masses then it must be able to access a new area of parameter space, or have substantially lower errors than previous methods. The later is difficult, as although the formal errors involved in this technique are small, many of the major systematic errors that dog other studies (e.g. in distances, inclinations etc) are still present, and likely dominate the total error budget. In this section we turn to demonstrating which areas of parameter space this technique can help explore. 

The obvious area of applicability for this technique is determining the variation of the M$_{\rm BH}$ - galaxy relations with Hubble type. 
Almost all spiral galaxies have molecular gas \citep[e.g.][]{1995ApJS...98..219Y}, and $\approx$22\% of ETGs have $\gtsimeq$10$^7$ \msun\ of H$_2$ \citep{2007MNRAS.377.1795C,Welch:2010in,2011MNRAS.414..940Y}. Not all these objects will have suitable molecular gas distributions, so the volume accessible with this technique needs to be large enough to allow us to obtain statistical samples.

In Figure \ref{almacarma} we show the minimum black hole mass detectable using this technique (at $\alpha$=5) as a function of distance, in a galaxy with V$_{\rm gal}=$100$\pm$5 \kms, using observations with 10 \kms channels and a resolution of 0\farc18 (the highest available with the CARMA telescope).  Also shown for reference are the galaxies with known black hole masses from the catalogue of \cite{2013ApJ...764..184M}, and the SMBH mass upper limits from \cite{2009ApJ...692..856B}. Approximately 80\% of the known objects could have their SMBH mass remeasured using this technique, and a similar fraction of the upper limits could be tested assuming all these objects had suitable molecular gas distributions. In reality galaxies with black hole masses estimated through stellar dynamics have usually been selected to ensure they do not have cold gas and dust, but this example simply shows the range of possible measurements  with respect to the current state of the field.
 
Figure \ref{almacarma2} shows the same as Figure \ref{almacarma}, but this time assuming the capabilities of full ALMA (a resolution of 0\farc011 at 345 GHz). When observing with ALMA essentially all the galaxies with known black-holes could in principle be re-observed. In addition we overplot in grey the distribution of all galaxies with central velocity dispersions listed in the Sloan Digital Sky Survey (SDSS) Data Release 7 catalogue \citep{2009ApJS..182..543A}, estimating their black hole masses with the best fitting M$_{\rm BH}$-$\sigma$ relation {for all galaxy types} from \cite{2013ApJ...764..184M}. Using ALMA, over 3.5$\times$10$^5$ of these objects ($\approx$45\% of the total) could in principle have their black-hole masses measured with $\alpha>5$. 
If higher observing frequencies and/or smaller channel sizes (combined with more accurate mass models) are used then this number would increase still further.
Even if only a small percentage of accessible galaxies have suitable molecular gas distributions, this technique has the potential to substantially increase the number of measured SMBH masses. We also highlight that (because of the behaviour of the angular diameter distance with redshift) with sufficiency sensitivity one could in principle measure the mass of a $\gtsimeq$4$\times$10$^8$ \msun\ black hole (with an inclination $>$30$^{\circ}$) at \textit{any} redshift.  

The above sections have shown that the molecular gas technique has great promise, however many challenges remain. 
The first of these hurdles is efficient target selection. In order to identify good targets one needs to determine that they have i) sufficient surface brightness in molecular gas to enable high resolution mapping ii) that the gas extends inwards near enough to the SMBH to make a measurement feasible iii) that the gas is kinematically relaxed and dynamically cold and iv) that it is possible to make a mass model of the luminous matter in the system at the required resolution. 

The first of these criteria means that one must select targets from existing single dish/low resolution interferometric surveys in order to estimate the molecular gas surface brightness (or conduct additional observations of likely target objects selected using other criteria).
The second and third criteria are hard to fulfil. Previous ionised gas surveys have found that selecting objects with regular, circular dust lanes that extend all the way to the galaxy centre can increase success rates for black hole mass measurements \citep{2002PASP..114..137H}, but if this holds in the same way for molecular gas has yet to be determined. The fourth point limits possible targets to those in which Hubble Space Telescope (or adaptive optics assisted infrared) imaging exists. In the future James Webb Space Telescope or very large ground based telescope data will be required to enable us to make mass models for objects further out in the universe. 

The final major challenge is dealing with non-circular motions that may be present in the molecular gas. This problem is present for all gaseous tracers, and the solutions developed for ionised gas measurements can also be used \citep[e.g.][]{2007ApJ...671.1329N}. Observing objects at intermediate inclinations would help ensure such non-circular motions can be identified, and provide sufficient information to constrain their effect and constrain extra degrees of freedom in the black hole mass fitting process.

\section{Conclusions}
\label{conclude}

In this work we have considered the strengths of using molecular gas kinematics to estimate black hole masses. We defined a figure of merit that will be useful in defining future observational campaigns, and discussed its implications. 
We showed that one can estimate black-hole masses even using data that only resolves scales $\approx$2 times the formal black hole sphere of influence, and confirm this be reanalysing lower resolution observations of the molecular gas around the black hole in NGC~4526. We also discussed the effect that velocity resolution and the depth of the galaxies potential have on the ability to estimate black hole masses. 

The next generation of very large optical telescopes (e.g. E-ELT, TMT, GMT) will, in principle, be able to reach reasonably similar angular resolutions to ALMA at 345 GHz (and thus explore similar areas of the black hole parameter space; \citealt{2014arXiv1401.7988D}). Once these large optical/infrared telescopes come online and ALMA has reached its full capabilities we will have an opportunity to enter a golden era in black-hole research, where we will be able to constrain the growth history of black holes, and thus their role in regulating galaxy formation.

\vspace{0.5cm}

\noindent \textbf{Acknowledgments}
The author thanks M. Bureau, E. Emsellem, R. van den Bosch and M. Sarzi for useful discussions, and the referee for suggestions that improved the paper. TAD also acknowledges the support provided by an ESO fellowship. The research leading to these results has received funding from the European
Community's Seventh Framework Programme (/FP7/2007-2013/) under grant agreement
No 229517. 

Funding for the SDSS and SDSS-II has been provided by the Alfred P. Sloan Foundation, the Participating Institutions, the National Science Foundation, the U.S. Department of Energy, the National Aeronautics and Space Administration, the Japanese Monbukagakusho, the Max Planck Society, and the Higher Education Funding Council for England. The SDSS Web Site is http://www.sdss.org/.
The SDSS is managed by the Astrophysical Research Consortium for the Participating Institutions. The Participating Institutions are the American Museum of Natural History, Astrophysical Institute Potsdam, University of Basel, University of Cambridge, Case Western Reserve University, University of Chicago, Drexel University, Fermilab, the Institute for Advanced Study, the Japan Participation Group, Johns Hopkins University, the Joint Institute for Nuclear Astrophysics, the Kavli Institute for Particle Astrophysics and Cosmology, the Korean Scientist Group, the Chinese Academy of Sciences (LAMOST), Los Alamos National Laboratory, the Max-Planck-Institute for Astronomy (MPIA), the Max-Planck-Institute for Astrophysics (MPA), New Mexico State University, Ohio State University, University of Pittsburgh, University of Portsmouth, Princeton University, the United States Naval Observatory, and the University of Washington.

\bsp
\bibliographystyle{mn2e}
\bibliography{bibFOM}
\bibdata{bibFOM}
\bibstyle{mn2e}
\label{lastpage}


\end{document}